\documentclass[aps,pra,twocolumn,superscriptaddress,10pt,article,nofootinbib,showpacs]{revtex4-1}
\usepackage{amsfonts}
\usepackage{amsmath}
\usepackage{amsthm}
\usepackage{braket}
\usepackage{graphicx}
\usepackage{hyperref}
\usepackage{color}
\usepackage{amssymb}
\usepackage{dsfont}
\usepackage{verbatim}
\usepackage{amsmath,amscd}
\usepackage{multirow}

\begin{document}

\title{Filling constraints on fermionic topological order in zero magnetic field}

\author{Nick Bultinck}
\affiliation{Department of Physics, Princeton University, Princeton, NJ 08544, USA}
\author{Meng Cheng}
\affiliation{Department of Physics, Yale University, New Haven, CT 06511-8499, USA}

\begin{abstract}
We consider two-dimensional electron systems in zero magnetic field at fractional filling. For such systems a Lieb-Schultz-Mattis theorem applies, forbidding the existence of a trivial insulator. However, the theorem does not distinguish between bosonic and fermionic systems. In this work we argue that in the case of fermionic systems, the topological orders that are compatible with the microscopic constraints are in general different from the bosonic case. We find different results in the case of even and odd denominator fillings, with even denominator fillings deviating stronger from the bosonic case. Part of our results also hold in three dimensions.
\end{abstract}

\maketitle

In their seminal work Lieb, Schultz and Mattis showed that a translationally invariant one-dimensional spin chain with half-integer spin per unit cell always has a vanishing energy gap if the Hamiltonian is invariant under spin rotations \cite{LSM}. This was the first instance of a result where microscopic properties were used to put general constraints on macroscopic properties of a quantum many-body system. Since then, this result has been extended in many different ways. It was shown by Oshikawa \cite{Oshikawa1}, and later more rigorously by Hastings \cite{Hastings}, how to generalize the Lieb-Schultz-Mattis argument to higher dimensions for systems with U$(1)$ symmetry. In all dimensions, the Lieb-Schultz-Mattis-Oshikawa-Hastings (LSMOH) argument excludes a unique ground state separated from the excited states by a non-vanishing gap. Above one dimension, however, this also allows for the possibility that the system develops topological order. Recent works have not only taken into account translation and internal U$(1)$ or SU$(2)$ symmetry to derive LSMOH constraints, but also different space group and on-site symmetries \cite{Parameswaran,Watanabe,Po}. More recently, it was shown that certain systems with a magnetic translation algebra can only preserve all the symmetries if they form a non-trivial symmetry-protected phase \cite{Lu,Yang}. Not only has the range of systems where LSMOH arguments apply been extended, there is also a conceptually new interpretation of the LSMOH-like theorems. In many cases, a system where LSMOH applies can be interpreted as the boundary of a higher dimensional gapped phase protected by both internal and space group symmetries via a bulk-boundary correspondance \cite{surface,Jian,Cho}.

The LSMOH arguments can not only be used to exclude a unique gapped symmetric ground state, they also constrain physical properties of the resulting symmetry-broken, gapless or topologically ordered phases. We will refer to these types of contraints as LSMOH constraints. For example, Oshikawa showed that Luttinger's theorem \cite{Luttinger} can be interpreted as a LSMOH constraint for Fermi liquids at non-integer filling \cite{Oshikawa2}. This was subsequently generalized in Refs. \cite{Vojta,Paramekanti,Bonderson}, where it was shown that if the system contains not only Landau quasi-particles but also topological excitations, Luttinger's theorem could be violated in very specific ways. An example of a system where this happens was dubbed the $FL^*$ phase \cite{SachdevS}. In this work we will also be interested in LSMOH contraints which follow from a fractional filling, but for systems that have no gapless excitations. 

For spin systems, the LSMOH contraints on a topologically ordered system are well-studied in recent years and by now well-understood using the framework of Symmetry-Enriched Topological (SET) phases \cite{Oshikawa4,zaletel,SET,Essin,Tarantino,surface,Qi}. For example, in systems with half integer spin per unit cell in two dimensions it is known that if there is an energy gap,  there must exist a quasiparticle excitation carrying half-integer spin (called the spinon), and another quasiparticle (the vison) which produces a minus sign after braiding with the spinon. An example of such a gapped state is a $\mathbb{Z}_2$ topological order, where the $\mathbb{Z}_2$ gauge charge/flux is the spinon/vison.
 Also strongly interacting bosons at half filling can develop an energy gap by forming a $\mathbb{Z}_2$ topological order \cite{Paramekanti,Isakov}. Since in these systems only translation symmetry and U$(1)$ spin z-component or particle number conservation play a role, the original LSMOH argument applies. The LSMOH argument does not distinguish whether the constituent particles are bosons or fermions, and excludes a trivial symmetric ground state in both cases. However, it is not clear if the LSMOH contraints have the same implications on the compatible topological orders. In fact, in this work we will argue that the LSMOH constraints do have different consequences for the topological order if the constituent particles are fermions. We do this by applying the algebraic theory of topological order in two dimensions and by invoking tensor network arguments, which hold in any dimension. At the end of the manuscript we also provide a consistency check with arguments based on vortex condensation. We find that even and odd denominator fillings behave differently, and that in many cases the minimal fermionic topological order will contain more superselection sectors than the minimal bosonic topological order at the same filling. But before going into these arguments, we first review the LSMOH contraints that arise as a result of fractional filling.

\begin{figure}
a) 
\includegraphics[width=0.09\textwidth]{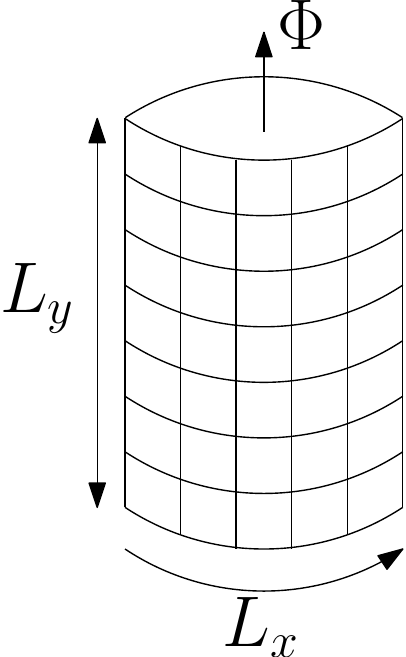} \hspace{0,5 cm}
b)
\includegraphics[width=0.17\textwidth]{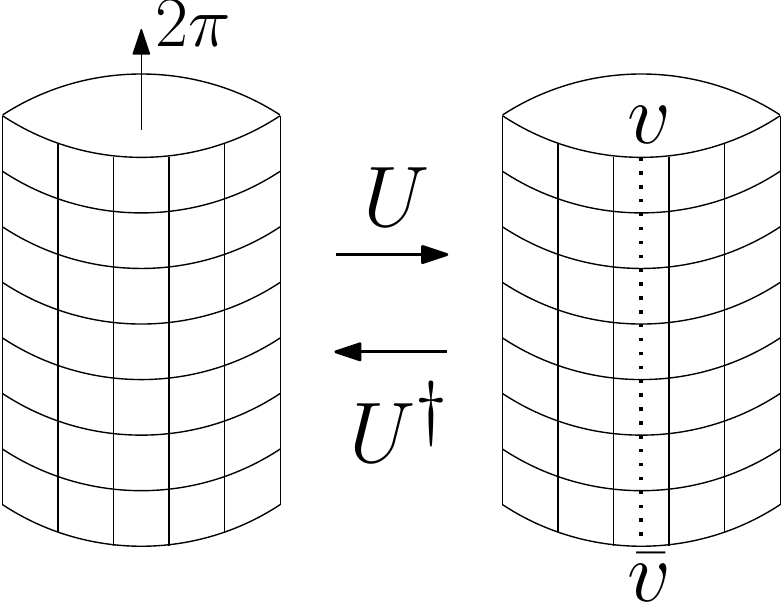}\caption{(a) A system on the cylinder with $L_xL_y$ unit cells. Flux $\Phi$ is inserted through the hole of the cylinder. (b) The large gauge transformation $U$ relates the state with $2\pi$ flux inserted to one with anyon flux $v$.}\label{cylinder}
\end{figure}

\emph{LSMOH contraints at fractional filling --} Consider a two-dimensional system on a cylinder with length $L_x$ along the periodic direction and length $L_y$ along the cylinder axis, such as shown in Fig. \ref{cylinder} (a). We define the filling as $\nu = N_e/L_xL_y$, where $N_e$ is the number of electrons and $L_xL_y$ the number of unit cells. We now follow Oshikawa and adiabatically insert a  flux $\Phi$ from 0 to $2\pi$ through the hole of the cylinder. We work in a gauge where $A_x=\frac{\Phi}{L_x}, A_y=0$. At the end of the adiabatic process, the Hamiltonian is related to the original Hamiltonian via a large gauge transformation, i.e. $H(2\pi) = U^\dagger H(0)U$. Here, $U$ is a unitary matrix of the form $U=\exp\left(\frac{2\pi i}{L_x} \sum_{\textbf{r}} x\, n_{\textbf{r}}\right)$, where the sum is over all unit cells, $x$ is the integer $x$-coordinate of the unit cell and $n_{\textbf{r}}$ is the particle number operator at unit cell $\textbf{r}$. The crucial observation is now that $U$ and $T_x$, the translation operator along the $x$-direction, satisfy following commutation relation: $T_x U = e^{2\pi i \nu L_y}UT_x$. So we see that at fractional filling and with the appropriate choice of $L_y$, the state obtained by adiabatically inserting $2\pi$ flux and subsequently applying the large gauge transformation has a different momentum than the state one started with. In this manuscript we are only interested in the scenario where the system is topologically ordered, so we will now explain how a system with anyonic excitations can accomodate for this momentum shift. In a topologically ordered system there is a ground state degeneracy on the cylinder, where each ground state is labeled by its anyon flux through the hole of the cylinder. This anyon flux can be measured by creating an anyonic excitation pair from the ground state, adiabatically transporting one anyon around the periodic direction of the cylinder and subsequently annihilating the pair again. After this process we again end up with the ground state, but we also pick up a phase which is called the braiding phase between the anyon flux and the transported anyon. These braiding phases unambiguously determine the trapped anyon flux. So let's assume that we started out with the state that has trivial anyon flux. After adiabaticaly evolving this state during the flux insertion process, we have to apply the unitary $U$ to bring us back to a ground state of the orginal Hamiltonian. However, here $U$ will map to a state that has an anyon flux $v$ through the hole of the cylinder. This is shown schematically in Fig. \ref{cylinder} (b). The momentum shift comes from the fact that there is an anyon $a$ background flux per unit cell, such that the braiding phase of $v$ with $a$ is $M_{a,v}=e^{2\pi i \nu}$ \cite{zaletel,surface}. To see why this is so, note that the effect of the background flux is that, although the string operator creating the $v-\bar{v}$ pair is invisible to local operators, it will nevertheless produce a phase when the path of the string operator is changed. Moving the string operator across a single site will precisely produce the braiding phase $M_{a,v}$. So if we translate the state on the cylinder with flux $v$ by one site along the $x$-direction, the string operator connecting $v$ and $\bar{v}$ will also get shifted by one site. Bringing this string operator back requires it to be moved across an entire column of sites, thus indeed producing the phase $M^{L_y}_{a,v}$, which matches the momentum shift obtained from microscopic arguments.

Because $v$ is obtained by inserting $2\pi$ flux, it follows from the non-trivial braiding phase $M_{a,v}$ that the background anyon $a$ carries a charge $\nu$ mod $1$, and can effectively screen the local particle density \cite{zaletel}. In fact we can make an even stronger statement. The requirement that the total charge carried by the background anyons has to be equal to the total charge carried by the microscopic particles fixes the charge of $a$ to be strictly equal to $\nu$, not modulo one. Up to now, all arguments applied to systems built from either bosons or fermions. However, the stronger statement that the charge of $a$ is strictly equal to $\nu$  is not relevant for bosonic systems (if there is no magnetic field \cite{Lee}), but does play an important role for fermionic systems.

In the following we will analyse what kind of topological orders are compatible with the LSMOH constraints if the constituent particles are fermions. We start by applying the algebraic theory of anyons in two dimensions, and treat the cases of even and odd denominator filling separately.

\emph{Even denominator filling --} Suppose the filling fraction is $\nu=\frac{p}{q}$ where $p$ and $q$ are coprime integers. We first consider the case where $q$ is even (so $p$ must be odd). We will assume that the fundamental fermions carry unit charge. From the LSMOH constraints we know that the topological order should have an Abelian anyon $a$ with U$(1)$ charge $\nu=p/q$. Fusing $q$ copies of the anyon $a$, we end up with a particle that has odd integer charge $p$. However, since the constituent particles are fermions, we know that every strictly local particle (i.e. a particle that braids trivially and has trivial topological spin) has even integer charge, thus $a^q\neq 1$. It was shown in Ref. \cite{Meng}, using the ribbon identity, that Abelian anyons in a fermionic topological order can always be written as $\mathcal{A}\times \{1,f\}$, where $\mathcal{A}$ is a braided tensor category\footnote{A (modular) braided tensor category is the mathematical framework that describes a bosonic topological order \cite{anyonsolved}.} and $f$ is the fundamental fermion (see also Ref. \cite{cano} for an earlier proof based on the $K$-matrix formalism). This excludes the possibility that $q$ copies of $a$ can fuse to the fundamental fermion. So the only remaining possibility is that $q$ copies of $a$ fuse to another anyon with odd integer charge, implying that the minimal fusion group generated by $a$ is $\mathbb{Z}_{2q}$.

An example of a topological order that meets the LSMOH requirements in the fermionic case is a $\mathbb{Z}_{2q}$ gauge theory, where the gauge charge has U$(1)$ charge $\nu$ and a large gauge transformation creates $2p$ copies of the fundamental gauge flux. This is to be compared with the bosonic case, where $a$ can be of order $q$, and a $\mathbb{Z}_q$ gauge theory can be realized.

Let's now focus on the physically relevant case of half filling, and consider the implications of the LSMOH constraints for a $\mathbb{Z}_4$ gauge theory as an example. We take the gauge charge $e$ to have U$(1)$ charge $1/2$, such that $e^3$ has charge $-1/2$ (mod $2$), and the gauge flux $m$ to be charge neutral. The opposite choice, with $e$ having charge $-1/2$ and $e^3$ having charge $1/2$ is equivalent since the theory has a topological symmetry $e\leftrightarrow e^3, m \leftrightarrow m^3$. The LSMOH constraints can now be satisfied if we take either $e$ or $e^3f$ as our background anyon $a$ (with the addition of possible gauge fluxes $m$), and if $v=m^2$ is created by a large gauge transformation. These options realize different SETs with translation and U$(1)$ symmetry.

\emph{Odd denominator filling --} We now consider the situation where $q$ is odd. In this case, there is no obstruction for $q$ copies of an anyon $a$ to fuse to the fundamental fermion. This is because one can simply redefine $a$ as $af$, such that now $q$ copies fuse to the trivial anyon. This of course does not come as a surprise since the topological orders described by the Laughlin states exactly have this property. However, the Laughlin topological orders also have a non-zero Hall conductance and are therefore not expected to occur in the systems at zero magnetic field without spontaneously breaking time-reversal symmetry.  

So let us turn to the case of discrete gauge theories, which can occur at zero magnetic field preserving time-reserval symmetry and are also relevant for our fermionic tensor network arguments later on. We first consider the case with odd $p$. Now $q$ copies of the background anyon $a$, which has charge $\nu=p/q$, fuse to a particle with odd integer charge. Assuming $a$ is one of the gauge charges, then in order for $q$ copies of $a$ to fuse into a fermion, we have to pick $a=e^kf$. The case with $k=1$ will then result in a $\mathbb{Z}_q$ gauge theory. For $p$ even, $q$ copies of $a$ can fuse to the trivial anyon if we take $a=e^k$, with suitable charge assignment on $e$. In the following section we show how the conclusions obtained above are consistent with the properties of fermionic tensor networks.

\begin{figure}
a)
\includegraphics[width=0.35\textwidth]{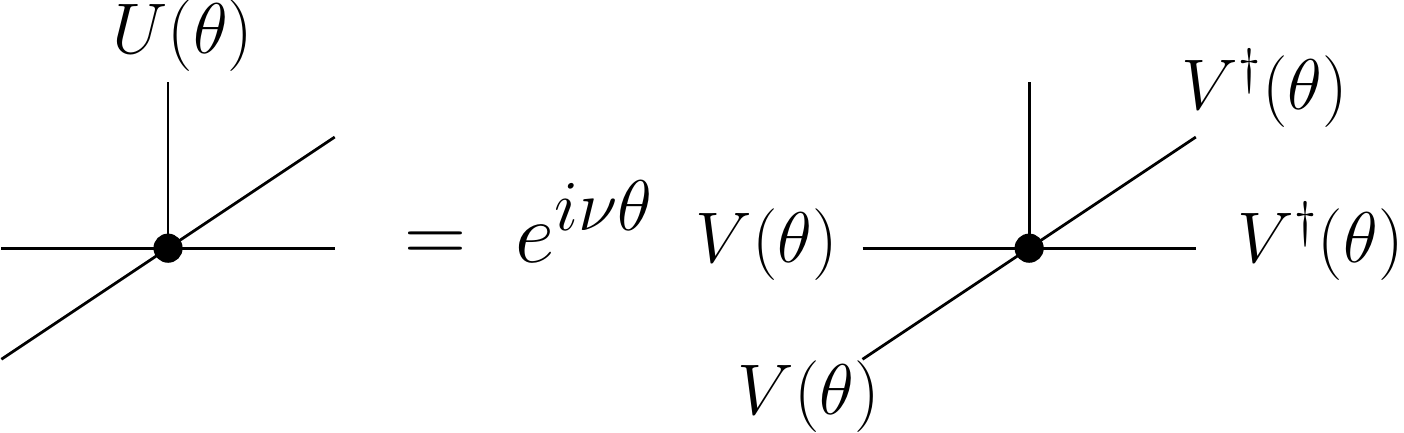} \\
b) 
\includegraphics[width=0.3\textwidth]{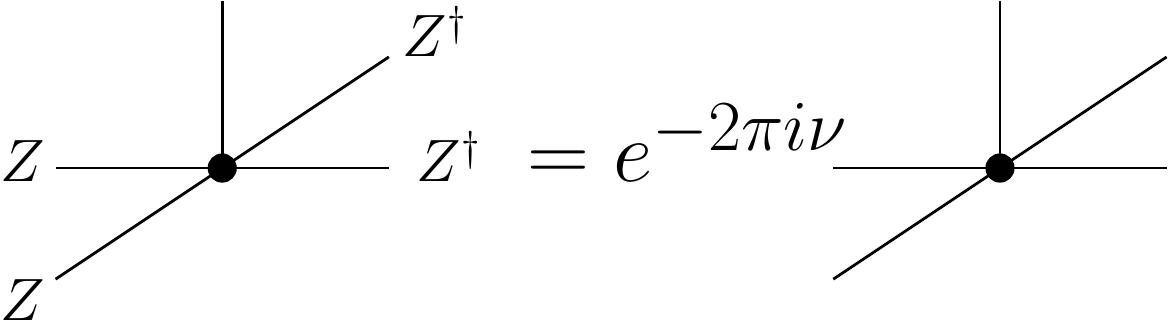}\caption{(a) Local tensor condition ensuring that the tensor network is at the required filling fraction. The vertical index, acted upon by $U(\theta)$, is the physical index. The other four indices are the virtual indices which are contracted with virtual indices of neighboring tensors in the formation of the tensor network. So this tensor would after contraction give rise to a two-dimensional tensor network on the square lattice. In the contraction process, the unitary matrices $V(\theta)$ cancel, such that the tensor network is an eigenstate of $U(\theta)^{\otimes L_x L_y}$ with eigenvalue $e^{i\nu L_x L_y \theta}$. (b) Tensor property following from evaluating (a) at $\theta=2\pi$. This purely virtual symmetry is the tensor network equivalent of the statement that there is topological order.}\label{bosonicTN}
\end{figure}

\emph{Fermionic tensor networks at fractional filling --} Let's first recall how topological order arises in bosonic tensor networks at fractional filling. A sufficient (but not necessary) condition for the local tensors that implements the U$(1)$ symmetry at fractional filling is shown in Fig. \ref{bosonicTN} (a). There $U(\theta)$ is the U$(1)$ action on the physical index, and $V(\theta)$ is a matrix acting on the virtual indices. The phase $e^{i\nu\theta}$ in Fig. \ref{bosonicTN} (a) ensures that the entire tensor network after contraction indeed has the right filling. We can now evaluate Fig. \ref{bosonicTN} (a) at $\theta=2\pi$, which gives the local tensor property in Fig. \ref{bosonicTN} (b), where we have defined $Z=V(2\pi)$ and used the fact that by definition $U(2\pi)=\mathds{1}$. Figure \ref{bosonicTN}(b) shows that the tensor is invariant, up to a phase $\gamma=e^{-2\pi i \nu}$, under the action of $Z$ on all virtual indices. This is exactly the tensor network equivalent of the statement that there is topological order with a background anyon per site \cite{Ginjectivity,Bultinck}. The minimal topological order compatible with this virtual symmetry is a $\mathbb{Z}_m$ gauge theory, such that $\gamma^m=1$. So writing the filling fraction as $\nu = p/q$, with $p$ and $q$ being coprime, the minimal topological order compatible with Fig. \ref{bosonicTN} (b) is that of a $\mathbb{Z}_q$ gauge theory \cite{Ginjectivity}. It is easy to see that a $\mathbb{Z}_q$ gauge theory is indeed compatible with the LSMOH constraints if the gauge charge carries $U(1)$ charge $\nu$ and the large gauge transformation creates $p$ copies of the fundamental gauge flux.

\begin{figure}
a)
\includegraphics[width=0.3\textwidth]{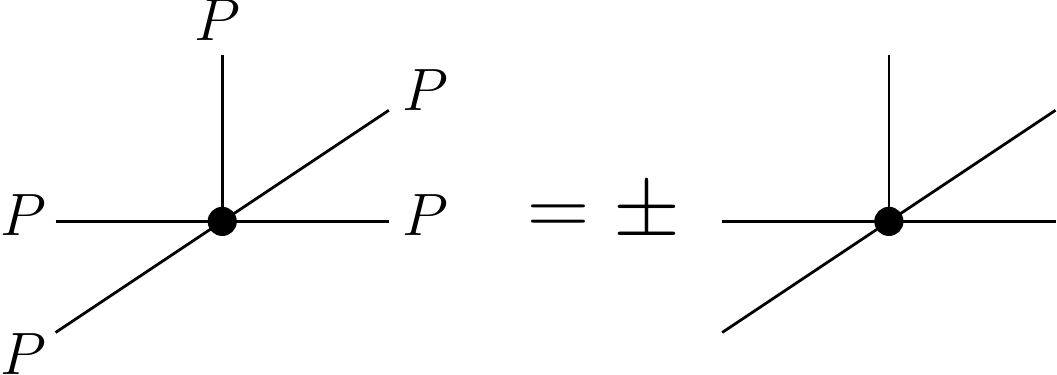} \\
b) 
\includegraphics[width=0.3\textwidth]{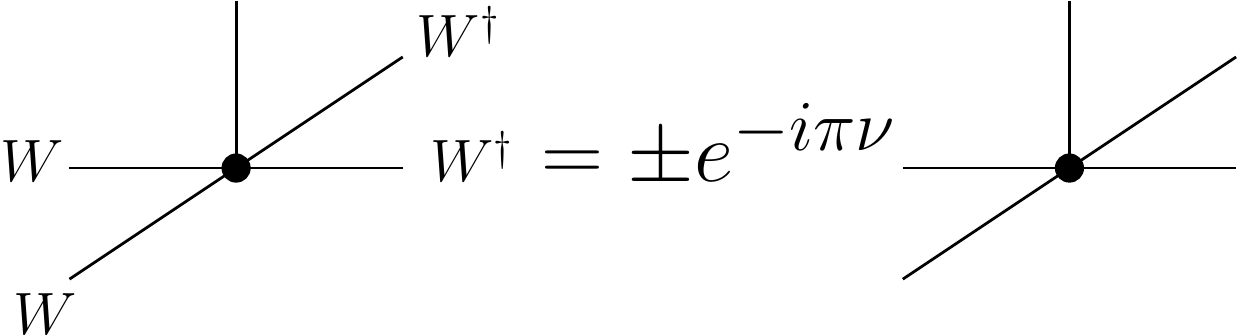}\caption{(a) Necessary condition for a fermionic tensor network to be well-defined: each tensor should have a well-defined fermion parity. The plus or minus sign denotes whether the tensor is even or odd. (b) Tensor property derived from (a) and Fig. \ref{bosonicTN}(a). We have defined $W\equiv PV(\pi)$.}\label{fermionicTN}
\end{figure}

Now let's turn to fermionic tensor networks. For fermionic tensor networks the local tensor properties of Fig. \ref{bosonicTN} can also be used to acquire the desired filling fraction. However, there is one additional requirement in the fermionic case, which is that the tensors should have a well-defined fermion parity. Otherwise, one can not make sense of a fermionic tensor network. This requirement means that there exists a fermion parity matrix $P$ for every index (we use the same symbol $P$, regardless of what index it acts on) satisfying $P^2=\mathds{1}$, such that the property in Fig. \ref{fermionicTN} (a) holds \cite{fMPS,fpeps}. Now crucially, it holds that $U(\pi)=P$ when acting on the physical index. From this we can derive the tensor property shown in Fig. \ref{fermionicTN} (b), where we have defined $W\equiv PV(\pi)$. Let's again denote the filling as $\nu=p/q$, with $p$ and $q$ coprime. Consider first the situation where $q$ is even. In that case, we find from the tensor identity in Fig. \ref{fermionicTN} (b) that the minimal topological order is a $\mathbb{Z}_{2q}$ gauge theory. For odd $q$, the situation depends on the fermion parity of the tensor and the value of $p$. For both an even parity tensor and odd $p$, or an odd parity tensor and even $p$ we again find that the minimal topological order is a $\mathbb{Z}_{2q}$ gauge theory. For the other cases, i.e. when the fermion parity of the tensor and $p$ are both even or odd, the minimal topological order is a $\mathbb{Z}_q$ gauge theory, just like for bosonic tensor networks. These results match with what we obtained previously based on the algebraic theory of anyons, if we adopt the natural intepretation for the minus sign in  Fig.\ref{fermionicTN} (a) as indicating whether the background anyon $a$ is of the type $e^k$ or $e^kf$. However, we note that although the figures were restricted to two-dimensional tensor networks, the results carry over to three and higher dimensional tensor networks without any modifications.

We can now also ask whether there are no topological orders being overlooked by the tensor network argument. In the case of half filling, the reasoning above based on the algebraic theory of anyons only shows that the minimal fusion group generated by $a$ should be $\mathbb{Z}_4$. This does not prove that the minimal topological order is a $\mathbb{Z}_4$ gauge theory. In fact, there is a natural candidate for a more minimal topological order to satisfy the LSMOH constraints, namely a fermionic $\mathbb{Z}_2$ gauge theory. However, this comes at the cost of breaking time reversal symmetry. From the group supercohomology construction \cite{Supercohomology, Gu}, we know that there are two fermionic $\mathbb{Z}_2$ gauge theories, which are mapped to each other under time reversal~\footnote{There are also ``beyond-supercohomology'' fermionic $\mathbb{Z}_2$ gauge theories, but they are incompatible with particle number conservation.}. Because the two fermionic $\mathbb{Z}_2$ gauge theories are related by time reversal, we can without loss of generality restrict ourselves to the one described by the $K$-matrix \cite{Gu}
\begin{equation}
K = \left(\begin{matrix}0 & 2 \\ 2 & 1 \end{matrix}\right)\sim \left(\begin{matrix}-4 & 0 \\ 0 & 1 \end{matrix}\right)\, ,
\end{equation}
where the equivalence relation is the usual $SL(2,\mathbb{Z})$ equivalence. We choose to work with the diagonal $K$-matrix. The fusion group of the fermionic $\mathbb{Z}_2$ gauge theory is $\mathbb{Z}_4\times\{1,f\}$, where the fundamental fermion is described by the vector $(0,1)^T$ and the generating anyon $a$ corresponds to the vector $(1,0)^T$. Imposing that the fundamental fermion has charge $1$ and the anyon $a$ charge $1/2$ to meet the LSMOH constraints fixes the charge vector to be $t=(-2,1)^T$. The fermionic $\mathbb{Z}_2$ gauge theory has a topological symmetry $a\leftrightarrow a^3$, such that the charge assignments of $1/2$ or $-1/2$ to $a$ correspond to equivalent U$(1)$ SETs. This topological order satisfies the LSMOH constraints if a $2\pi$ flux insertion creates the anyon $a^2f$. This gives the correct braiding phase since $M_{a,a^2}=-1$ and $f$ is transparant. Only $a^2 f$ can bind to a $2\pi$ flux since it has charge zero modulo $2$, and is therefore consistent with the Hall conductance, which is zero. The reason this topological order is overlooked by the tensor network analysis is that a fermionic $\mathbb{Z}_2$ gauge theory requires a non-trivial matrix product operator action on the virtual indices \cite{fpeps}. We would like to note that although the fermionic $\mathbb{Z}_2$ gauge theory passes all the LSMOH checks, and has a natural explanation why it is missed by our tensor networks, we can not guarantee that this phase can indeed be realized by fermions at half-filling. We also want to remind the reader that for odd $q$, there are no fermionic $\mathbb{Z}_q$ gauge theories, as follows from supercohomology \cite{BiYou}. 

\emph{Vortex condensation --} As a final argument, we show that the results above are consistent with a heuristic reasoning based on vortex condensation. For simplicity, we restrict to the case of half filling. For bosonic systems it is known that in the superfluid phase the filling fixes the phase acquired by a vortex when it moves around a unit cell to be $\pi$ \cite{Paramekanti}. This of course also follows from particle-vortex duality \cite{Dasgupta,Fisher,Peskin,Stone}. So if we want to destroy the superfluid by condensing the vortices, we can only condense $4\pi$ vortices if we are to preserve translation symmetry. The $2\pi$ vortices survive as topological excitations \cite{Senthil}, accompanied by a charge $1/2$ boson (which is required for the topological order to be modular), leading to a $\mathbb{Z}_2$ gauge theory for the gapped symmetric phase.

In the fermionic case we have a pair condensate such that vortices with vorticity $k$ bind $k\pi$ flux. A similar argument as in the bosonic case shows that now the minimal vortex we can condense without breaking translation symmetry has vorticity $k = 4$ \cite{Shankar,Balents}. After the condensation, the gapped symmetric phase will have three topological excitations with trivial topological spins and charges $m \frac{2}{k}=m/2$ ($m=1,2,3)$, and three vortices surviving as charge neutral topological excitations \cite{Metlitski1,Potter,Wang,Metlitski2}.  Note that we did not discuss the statistics of the vortices, i.e. it is possible that the vortices are not bosons, or even non-Abelian when $k$ is odd \cite{Volovik,Read}. But in the simplest case the vortices will be bosons and after the condensation we end up with a $\mathbb{Z}_4$ gauge theory.

\emph{Conclusions --} We have studied the implications of the LSMOH constraints arising from fractional filling on the gapped symmetric phases in fermionic systems. We found that although the LSMOH theorem does not distinguish between bosonic and fermionic systems, the symmetric gapped phases at fractional filling are affected in a different way depending on the nature of the constituent particles. Especially for even denominator filling, the fermionic topological order can not simply be the minimal topological order occuring in the bosonic case, i.e. a $\mathbb{Z}_q$ gauge theory. 

We have considered systems in zero magnetic field, but it would be interesting to extend the current analysis to lattice models with non-zero flux per plaquette. It was shown in Ref. \cite{Oshikawa3} that in the presence of a magnetic field the microscopic properties can also constrain the value of the Hall conductance. Again, all arguments in Ref. \cite{Oshikawa3} apply to both systems consisting of bosons or fermions. However, the constraints on the Hall conductance will have different implications depending on the nature of the constituent particles. For example, it is known that in bosonic systems the minimal value for the Hall conductance compatible with the absence of non-trivial anyons is $2$ \cite{Chen,Vishwanath,Levin}, while in the fermionic case it is $1$.  

It would also be worth trying to develop the tensor network methods further. In Ref. \cite{Williamson}, a general framework for SET phases with discrete on-site symmetries in bosonic tensor networks was developed. It was also shown in Refs. \cite{Jian1,Jian2} how spatial symmetries can systematically be taken into account. Extending these formalisms to fermionic tensor networks would provide an independent derivation of all the (non-chiral) topological orders that are compatible with a particular LSMOH constraint.

\emph{Acknowledgements --} NB would like to thank Shinsei Ryu and Michael Zaletel for inspiring discussions. NB is supported by a BAEF Francqui Fellowship. MC is supported by startup funds from the Yale University.

\bibliography{Fillingconstraints}

\end{document}